\documentclass[aps,pra,twocolumn,showpacs,floatfix,amsmath,amssymb,unsortedaddress]{revtex4-1}

\usepackage{graphicx} 
\usepackage{bm} 
\usepackage[utf8x]{inputenc}
\usepackage{times}
\usepackage{flafter} 
\usepackage{textgreek} 
\usepackage{mathrsfs} 

\begin{document}

\def\vekt#1{\textbf{#1}}
\def\vektp{\vekt{p}}
\def\vektA{\vekt{A}}
\def\vektE{\vekt{E}}
\def\vekte{\vekt{e}}
\def\vektr{\vekt{r}}
\def\Ip{I_{\mathrm{p}}}
\def\Up{U_{\mathrm{p}}}
\def\diff{\mathrm{d}}
\def\ncyc{n_\mathrm{c}}

\title{Two-color phase-of-the-phase spectroscopy in the multiphoton regime}

\author{M.~A.~Almajid, M.~Zabel}
\author{S.~Skruszewicz}
\altaffiliation{Present address: Institut f\"ur Optik und Quantenelektronik, Friedrich-Schiller-Universit\"at Jena, Max-Wien-Platz 1, 07743 Jena, Germany}
\author{J.~Tiggesb\"aumker}
\author{D.~Bauer}
\thanks{Corresponding author: dieter.bauer@uni-rostock.de}
\affiliation{Institut f\"ur Physik, Universit\"at Rostock, 18051 Rostock, Germany}

\date{\today}
\begin{abstract} 
  Momentum-resolved photoelectron emission from xenon in colinearly polarized  two-color laser fields at above-threshold ionization
  conditions is studied both
  experimentally and theoretically.  We utilize phase-of-the-phase spectroscopy   as recently introduced by Skruszewicz {\em et al.}, Phys. Rev. Lett. 115, 043001 (2015) to analyze the dependence of the yields on the relative
  phase $\varphi$ between the fundamental and second harmonic laser
  fields. The resulting phase-of-phase spectra feature a characteristic checkerboard
  pattern, which can analytically be described within the strong-field
  approximation.

\end{abstract} 

\maketitle

\section{Introduction} The development of lasers has revitalized
atomic, molecular, chemical, and optical physics.  The interaction of
intense laser pulses with matter gives rise to a wealth of phenomena that are nonlinear in both the laser intensity and the number of photons involved, examples being multiphoton ionization (MPI), above-threshold
ionization (ATI), or tunneling ionization (see, e.g.,
\cite{Milosevic2006} for a review). While for weak laser fields MPI
could still be tackled in a perturbative fashion where each
additionally absorbed photon requires the next-order, conventional
perturbation theory becomes inadequate for stronger fields due to the
lack of a small parameter and very pronounced AC-Stark
effects \cite{delone2000multiphoton}. Instead, the so-called strong-field approximation (SFA) (see, e.g., 
\cite{Milosevic2006,Popruzhenko2014a} for a recent review) became the work horse of
choice and particularly insightful when interpreted in terms of
quantum orbits \cite{Milosevic2006,Popruzhenko2014a}. For instance, ATI peaks
separated in energy by $\hbar\omega$ can be understood as the
interference of quantum orbits of electrons emitted in subsequent
laser cycles (inter-cycle interference) while interfereing quantum
orbits originating from the same laser cycle (intra-cycle
interference) lead to other features in photoelectron spectra (PES) \cite{Arbo10},
such as holographic side lobes \cite{Huismans2011,huismansprl2012}.

Patterns in the photoelectron spectra clearly depend on the laser
pulse and the target. The dependency on the particulars of the laser
field allow for the control of the ionization dynamics or the precise
experimental characterization of the laser pulse \cite{chini2014}. The dependency on
the target opens up the possibility to image the system employing its
own electrons \cite{Blaga2012}.

In this work, we do not consider the PES themselves but the {\em
change} of the momentum-resolved yield as function of a the relative phase $\varphi$ between the
$\omega$ and the $2\omega$ component of a colinearly polarized
two-color pulse with vector potential (in dipole approximation)
\begin{equation}\label{vp} \textbf{A}(t) = A_0(t)\Big[\sin \omega t +
\xi \sin(2 \omega t + \varphi) \Big]{\bf e}_{z},
\end{equation} where $\xi$ is a relative field amplitude, and $A_0(t)$
is the envelope of the laser pulse. Note that if the ratio of $2\omega$ to $\omega$ component of the vector potential is $\xi$, the ratio of the corresponding electric field components is $2\xi$ because $\vektE=-\partial_t\vektA$. The ratio in intensities then is $(2\xi)^2$. Below we will consider $\xi=0.05$, i.e., the $2\omega$ component of the electric field is 10\% of the fundamental, and the intensity ratio is 1\%.

A plethora of two-color studies have been performed recently, with different combinations of laser frequencies and polarizations 
\cite{Gong2017,EickeLein2017,Busulad2017,Vvedenski2017,Zhang2017,Hamilton2017,Song2017,Natan:16,Li:16,Yu2016,mancuso2016,richter2016,Das2016,petersson2016,Zheng2015,Skrusz2015,ArboKitzler2014}. 
Important foreseeable applications of two-color fields are the efficient generation of THz radiation from plasmas \cite{Zhang2017} or the control of strong-field phenomena by steering the electron emission \cite{ZhoOE11}, for instance. Atoms and molecules in two-color fields are also interesting in themselves as they allow a deeper understanding of the ionization dynamics due to the high sensitivity of strong-field observables to the detailed shape of the electric field, which is modified by the relative phase $\varphi$  on a sub-cycle time scale. Relative-phase dependent measurements are thus  a more challenging, stringent test for theory and, perhaps most
importantly, offer the possibility to distinguish coherent and incoherent
contributions to PES or other observables. Here, ``coherent'' and ``incoherent'' mean that
the ionization dynamics that lead to photoelectrons with a certain
final momentum in PES do depend or do not depend on the precise
shape of the electric field of the laser, and thus on the relative
phase $\varphi$. The subcycle ionization dynamics in few-active
electron systems such as atoms or small molecules are expected to be
coherent in this sense while thermal enmission from larger systems is
expected to be incoherent. Before we aim at complex targets though, we
need to understand the canonical, ``textbook'' phase-of-the-phase (PP) spectra originating
from coherent electron emission. In Ref.~\cite{Skrusz2015}, phase-of-phase spectra
were found to show a typical structure of two overlapping clubs due to
rescattering in strong laser fields. In this work we investigate phase-of-phase spectra at lower laser
intensities, i.e., in the multiphoton ionization (MPI) regime.

The paper is organized as follows:  The fundamental idea behind phase-of-the-phase spectroscopy is introduced and a generic experimental result for Xe is presented in Sec.~\ref{sec:experiment}. Simulation results and the derivation of the ionization rate for a two-color pulse, both based on the SFA, are presented in Sec.~\ref{theory}, followed by a discussion in Sec.~\ref{disc}. We briefly summarize and give an outlook in Sec.~\ref{concl}. 

Atomic units are used unless
indicated otherwise. The dipole approximation is valid for the laser
parameters under study and applied throughout.

\section{Experimental result and phase-of-the-phase analysis} \label{sec:experiment}
Details of the experimental setup to measure phase-sensitive signals
from two-color laser pulse excitations of atoms and molecules has been
described previously \cite{Skrusz2015}. Briefly, a Ti:sapphire laser
system produces near-infrared pulses of energy $E_\mathrm{L} =2.5$\,mJ and
$\lambda=794$\,nm with a pulse duration of 180\,fs at a repetition
rate of 1\,kHz. To sculpture the pulses on the time scale of the
electric field oscillation the fundamental is superimposed by a weak
second harmonic field with an intensity ratio $I_{2\omega}/I_\omega=0.01$. The temporal
overlap between fundamental and second harmonic is controlled by birefringent
calcite crystals and two glass wedges. Additional wave plates are used
to align the polarization axes parallel to each other.  The wedges
are mounted on piezo-driven motors to achieve sub-cycle control over
the relative phase $\varphi$ between $\omega$ and $2\omega$ component. A
concave silver-mirror with a focal length of 250\,mm focuses the
pulses into the interaction region of a homebuilt velocity map imaging (VMI)
electron spectrometer~\cite{SkrRSI14}, giving pulse intensities of
$I_\omega \simeq 5\times 10^{13}$\,W\,cm$^{-2}$. Rare gases enter as an
effusive beam from a nearby capillary.

\begin{figure}
    \includegraphics[width=0.75\columnwidth]{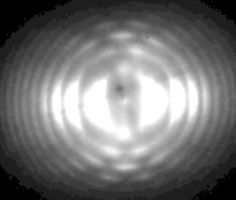}
    \caption{Screenshot of the VMI detector during a colinearly
polarized two-color experiment with Xe.  Laser
polarization in horizontal direction. Laser parameters are given in the text. ATI rings separated by
$\hbar\omega$ in energy and the ``carpet'' structure \cite{Korneev2012} are clearly visible.  }
    \label{fig:VMIsnapshot}
\end{figure}

Figure~\ref{fig:VMIsnapshot} shows a typical snapshot of the VMI
screen for such a two-color experiment on xenon atoms.  
ATI rings and the previously studied ``carpet'' structure
\cite{Korneev2012} are clearly visible. However, because the $2\omega$
component is just a weak perturbation, no big changes are seen on the
VMI screen with the naked eye when the relative phase $\varphi$ is
changed. 
 The idea behind phase-of-the-phase spectroscopy is to study systematically the {\em
change} in the photoelectron yield as a function of the relative phase
$\varphi$ (or any other periodic parameter, e.g., the carrier-envelope
phase \cite{Zhereb2012}).  Assuming that the momentum-resolved yield
$Y(\vektp,\varphi)$ behaves predominantly as
\begin{equation} Y(\vektp,\varphi) \simeq Y_0(\vektp) + \Delta
Y_1(\vektp) \cos[\varphi+\Phi_{1}(\vektp)] \label{yield}
\end{equation} for most final photoelectron momenta $\vektp$, the change in the
yield can be characterized by just two functions of $\vektp$: the relative phase
contrast (RPC) $\Delta Y_1(\vektp)\geq 0$ and the phase of the phase (PP) $\Phi_1(\vektp) \in [-\pi,\pi)$ that
tells us whether the yield changes, e.g., $\pm\cos$-like with
$\varphi$ or $\pm\sin$-like. Technically, phase of phase and relative phase contrast can be quickly
determined by fast-Fourier-transforming $Y(\vektp,\varphi)$ with
respect to $\varphi$ for each $\vektp$. The yield may
actually change with twice the relative phase for certain final
momenta $\vektp$, i.e., $\sim\Delta Y_2(\vektp) \cos[2\varphi+\Phi_2(\vektp)]$. In fact, this is
the case in leading order $O(\xi^2)$ for electron emission
perpendicular to the polarization direction because $\Delta
Y_1=0$ there.  However, in this work we limit our discussions to
the fundamental PP $\Phi_1$, i.e., to order $O(\xi)$, as will be shown in
Section~\ref{theory} below.

Figure~\ref{fig:exp:RPCandPP} shows the experimental RPC and PP
spectra for xenon, determined from a series of snapshots for varied relative phase $\varphi$ like the one
in Fig.~\ref{fig:VMIsnapshot}. For a fixed relative phase the average of $10000$ PES is taken. The relative phase is sampled in steps of   $\simeq 2\pi/70$.  A checkerboard pattern in the PP along the ATI rings is clearly visible.
The color coding can be cyclically shifted because only the change in
the relative phase is controlled in the experiment, not its absolute
value. We chose the color-coding such that the checkerboard pattern is
red and blue, representing $\pm\sin$-like changes in the yield
$Y(\vektp,\varphi)$ according to the pulse form \eqref{yield} assumed in
the theoretical analysis.

\begin{figure}
    \includegraphics[width=0.8\columnwidth]{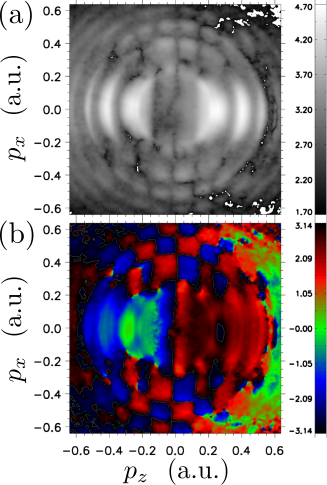}
    \caption{Experimental RPC $\Delta Y_1(\vektp)$ (a) and PP
$\Phi_{1}(\vektp)$ (b) spectra for Xe. Parameters as for
Fig.~\ref{fig:VMIsnapshot}, given in the text. A checkerboard pattern oriented along the ATI rings is visible in the PP spectrum (b).}
    \label{fig:exp:RPCandPP}
\end{figure}

\section{Theory} \label{theory} 
\begin{figure}
    \includegraphics[width=0.8\columnwidth]{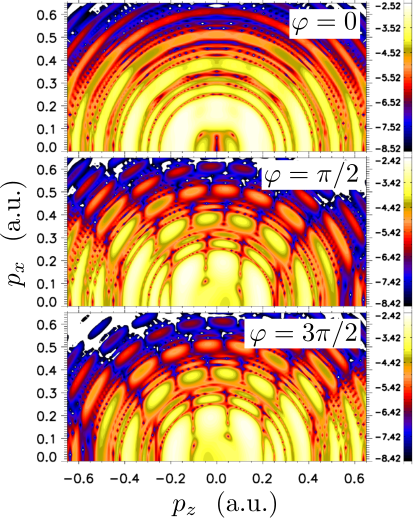}
    \caption{ Logarithmically color-coded SFA-PES for $\Ip=0.445$ (Xe), vector potential amplitude
$A_0=2/3$, $\omega=0.057$, $20$-cycle $\sin^2$ pulse, $\xi=0.05$,
and $\varphi=0$, $\pi/2$, $3\pi/2$, calculated from
\eqref{direct}.} \label{fig:sfa:matrixelem}
 \end{figure}

We apply the SFA for the so-called direct
electrons, i.e., those that do not rescatter at the parent ion.  
In length gauge, the SFA matrix element reads \cite{Milosevic2006,Popruzhenko2014a} \begin
{equation}\label{direct} M^{(\text{SFA})}_{\textbf{p}} = -\mathrm{i}
\int^{\infty}_{-\infty}\langle
\Psi^{\text{GV}}_{\textbf{p}}(t)|\textbf{r}\cdot
\textbf{E}(t)|\Psi_{0}(t)\rangle \, \diff t
\end {equation} where
\begin {equation}\label{volkovstate}
|\Psi^{\text{GV}}_{\textbf{p}}(t)\rangle=
\mathrm{e}^{-\mathrm{i}S_{\textbf{p}}(t)}|\textbf{p}+
\textbf{A}(t)\rangle
\end {equation} is a Gordon-Volkov (GV) state, i.e., a solution to the
time-dependent Schr\"odinger equation
$\mathrm{i}|\dot{\Psi}^{\text{GV}}_{\textbf{p}}(t)\rangle =
[\vektp^2/2 +
\vektr\cdot\vektE(t)]|\Psi^{\text{GV}}_{\textbf{p}}(t)\rangle$ without
binding potential but laser only, $|\Psi_{0}(t)\rangle =
\exp(\mathrm{i}\Ip t) |\Psi_{0}(0)\rangle$ is the initial bound state,
and
\begin{equation} S_{\textbf{p}}(t)=\frac{1}{2} \int^{t}[\textbf{p}+
\textbf{A}(t')]^{2} \, \diff t'
\end{equation} is the Coulomb-free, classical action. In dipole approximation, the PES $|
M^{(\text{SFA})}_{\textbf{p}}|^2$ is azimuthally symmetric about the
laser polarization axis in $\vekte_z$ direction. For illustration, PES
calculated with the SFA for laser intensity $I = 5 \times
10^{13}\,$W/cm$^2$, $\lambda=800$\,nm ($\omega=0.057$), $20$-cycle $\sin^2$
pulse, $\xi=0.05$, relative phases $\varphi=0,\pi/2,3\pi/2$, and
$\Ip=0.445$ (Xe) are shown in Fig.~\ref{fig:sfa:matrixelem}. The
initial state $|\Psi_{0}(0)\rangle$ affects the result only as a
preexponential form factor $\langle \vektp + \vektA | \vektr
|\Psi_{0}(0)\rangle \cdot \vektE(t)$ \cite{Milosevic2006,Popruzhenko2014a}. For simplicity, we chose a hydrogen-like 1s state for
$|\Psi_{0}(0)\rangle$ with $\Ip$ adjusted to Xe.
  For all $\varphi$ ATI rings are clearly visible. For $\varphi=0$
(or $\pi$), the SFA-PES is left-right symmetric (i.e., invariant under
the transformation $p_z \to -p_z$). Instead, for $\varphi=\pi/2$ and
$3\pi/2$ the weak $2\omega$ component of the laser field introduces a
visible left-right asymmetry: the ATI peaks appear to be more pronounced and shifted along
the ATI rings. These shifts could be reproduced extending the quantum orbit approach and the calculation of the curves of destructive interference as outlined in \cite{Korneev2012} to two-color fields. However, we aim at a closed analytic expression for the ionization probability as a function of the relative phase in this work and thus follow a different route.

\begin{figure}
    \includegraphics[width=0.8\columnwidth]{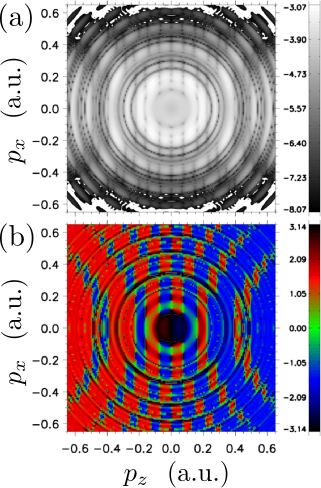}
    \caption{RPC (a) and PP (b) calculated from PES for 20 values of  $\varphi\in
[0,2\pi)$ and the parameters of
Fig.~\ref{fig:sfa:matrixelem}. } \label{fig:sfa:matrixelemRPCandPP}
 \end{figure}

Figure~\ref{fig:sfa:matrixelemRPCandPP} shows the RPC $\Delta Y_1(\vektp)$  and the PP $\Phi_1(\vektp)$
 for the parameters of
Fig.~\ref{fig:sfa:matrixelem}. A checkerboard pattern similar to the one in the experimental PP spectrum in   Fig.~\ref{fig:exp:RPCandPP}b is observed, at least for sufficiently big lateral momenta $|p_x|$. The PP signature for small momenta is different because of the neglect of Coulomb effects on the emitted electrons in the SFA \cite{Bauer2006a,Popruzhenko2014a}. The fact that there is a checkerboard pattern in PP spectra at all is simple to understand: if an ATI
peak moves along its ATI ring, the yield increases in regions where
previously was less probability and decreases at the initial location. As a consequence, only two discrete PPs show up. Why the yield within an SFA
treatment behaves $\pm\sin\varphi$-like and not with some other pair
of PPs is less obvious but will be derived analytically in the following.

\begin{figure}
    \includegraphics[width=0.8\columnwidth]{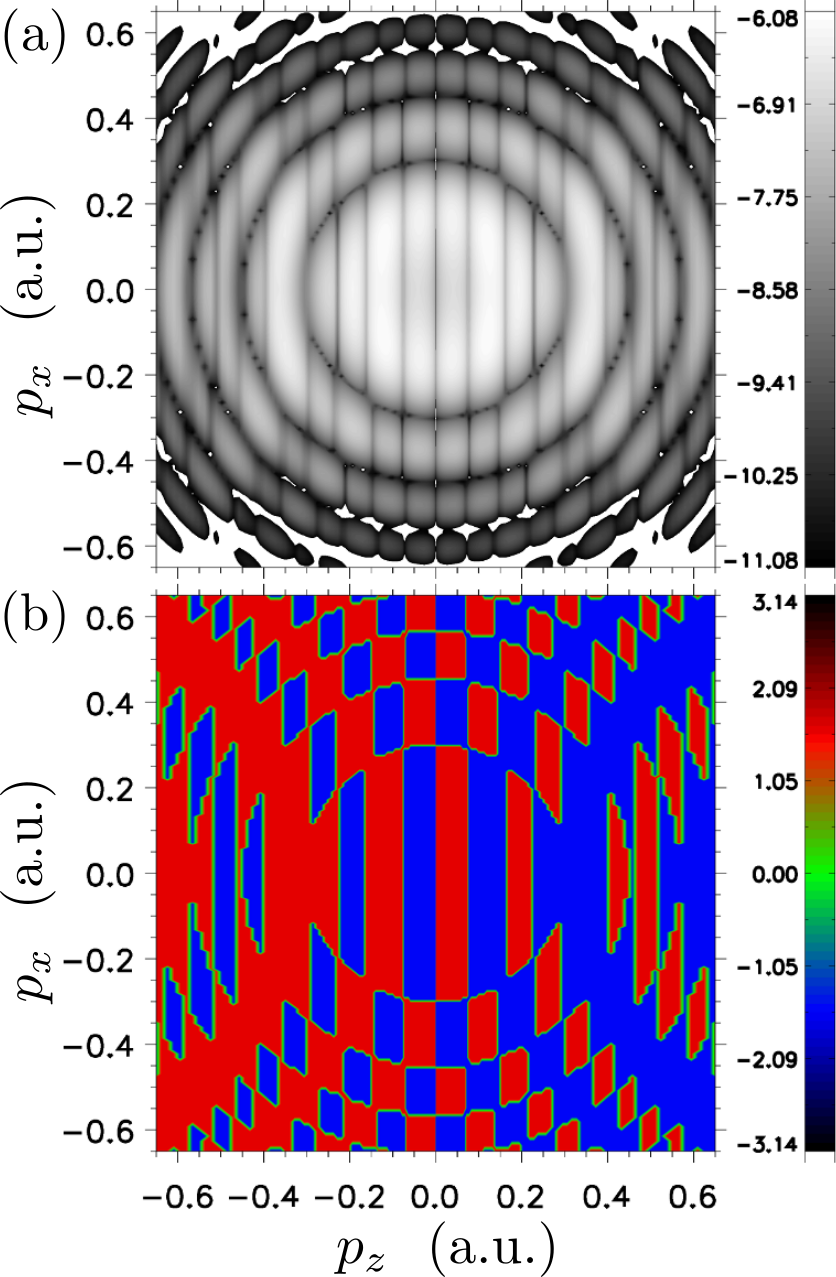}
    \caption{RPC (a) and PP (b) spectra calculated using the flat-top,
infinite pulse SFA result for the rate Eq.~\eqref{bessel2}.}
    \label{fig:sfa:besselian}
\end{figure}

In order to obtain an analytical result we work in velocity gauge and
consider long flat-top pulses of vector potential amplitude $A_0$ that are switched on and off
at $\mp\infty$.  It is known that the SFA is plagued by gauge
non-invariance, leading, in general, to different PES for velocity
gauge (with $\vektp\cdot\vektA(t)$ as the interaction term) and length
gauge (with $\vektr\cdot\vektE(t)$, with the electric field
$\vektE=-\partial_t\vektA$) \cite{Bauer2005}. However, in this work we use an
$1$s-state as the inital state for the SFA and tune the
ionization potential $\Ip$ to the experimental value. In that case
there is no visible difference between length and velocity gauge PES obtained numerically from the SFA.
The SFA matrix element relevant for the ionization rate
$\Gamma_\vektp$ can then be written in the form \cite{Reiss1980}
\begin{equation}\label{mvelocity} {M'}^{(\text{SFA})}_{\textbf{p}} =
\mathrm{i}\Psi_{0}(\textbf{p})\left( \frac{p^{2}}{2} + \Ip\right)
\int^{\infty}_{- \infty} \diff
t~\mathrm{e}^{\mathrm{i}S_{\textbf{p},\Ip}(t)},
\end{equation} with $\Psi_{0}(\textbf{p}) = \langle
\textbf{p}|\Psi_{0}\rangle$ and
\begin{equation}\label{s} S_{\textbf{p},\Ip}(t)=
\int_{-\infty}^{t}\left\{\frac{1}{2}[\textbf{p}+ \textbf{A}(t')]^{2} +
\Ip \right\}\, \diff t'.
\end{equation} Inserting \eqref{vp} with $A_0(t)=A_0$ into \eqref{s},
\begin{align}\label{ss} S_{\textbf{p},\Ip}(t) = &
\left(\frac{p^{2}}{2} + \Ip + \Up\right)t\\ & - \frac{A_0
p_z}{\omega}\sin\left(\omega t+{\pi}/{2}\right) + \frac{A_0^2}{8
\omega} \sin\big[2(\omega t +\pi/2)\big] \nonumber \\ &
-\frac{A_0\xi}{\omega}\bigg[\frac{p_z}{2} \cos(2\omega t + \varphi) +
\frac{A_0}{6} \sin(3\omega t + \varphi) \nonumber \\ & \qquad\qquad -
\frac{A_0}{2} \sin(\omega t + \varphi)\bigg] + {\cal{O}}(\xi^2)
\nonumber
\end{align} is obtained. Here, $\Up=A_0^2/4$ is the ponderomotive energy.  With the help of the generalized Bessel functions
$J_{n}(u, v)$,
\begin {align} \mathrm{e}^{\mathrm{i} [ u\sin \phi + v \sin(2\phi)]} =
& \sum^{\infty}_{n=-\infty} \mathrm{e}^{\mathrm{i}n\phi}J_{n}(u, v),
\\ J_n(u,-v) = & (-1)^n J_{-n}(u,v),
\end{align} we can evaluate the matrix element \eqref{mvelocity},
neglecting terms of order ${\cal{O}}(\xi^2)$, as
\begin{align}\label{matrix} {M'}^{(\text{SFA})}_{\textbf{p}} = & 2\pi
\mathrm{i} \Psi_{0}(\textbf{p})\bigg(\frac{p^2}{2} + I_p\bigg) \\ &
\times \sum^{\infty}_{n=-\infty}\mathrm{i}^{n}
\delta\Big(\frac{p^2}{2} + \Ip + \Up - n\omega \Big) \Bigg\{ J_{n}
\nonumber \\ & + \frac{\mathrm{i}A_0 \xi}{4\omega}\bigg[\Big(A_0
J_{n+1}+ p_zJ_{n+2} + \frac{A_0}{3} J_{n+3}\Big)
\mathrm{e}^{\mathrm{i}\varphi} \nonumber \\ & + \Big( A_0J_{n-1} +
p_zJ_{n-2} +\frac{A_0}{3} J_{n-3}\Big) \mathrm{e}^{-
\mathrm{i}\varphi} \bigg] \Bigg\} . \nonumber
\end{align} Here we suppressed for brevity the arguments of all the generalized
Bessel functions, i.e., \begin{equation} J_n=J_n(u,v), \quad u=-\frac{p_zA_0}{\omega}, \quad v= -\frac{\Up}{2\omega}. \end{equation} 
For the rate follows, analogously to the one-color calculation in, e.g., Refs.~\cite{Reiss1980,Mulser2010},
\begin{align}\label{bessel} \Gamma_\vektp = &
2\pi|\psi_{0}(\textbf{p})|^{2}\left(\frac{p^2}{2}+ \Ip\right)^{2} \\ &
\times \sum^{\infty}_{n=-\infty} \delta \Big(\frac{p^2}{2} + \Ip + \Up
- n \omega \Big) \Bigg\{ J_{n}^{2} - J_{n}\xi \sin \varphi \nonumber
\\ & \times \Bigg[4 v( J_{n-1} - J_{n+1}) + \frac{u}{2} (J_{n-2} -
J_{n+2} ) \nonumber \\ & + \frac{4v}{3}(J_{n-3} - J_{n+3} ) \Bigg]
\Bigg\}.
\end{align} We can eliminate the terms $\sim J_{n\pm 3}$ using the property of the generalized Bessel functions $2nJ_n
= u(J_{n-1}+J_{n+1}) + 2v (J_{n-2} - J_{n+2})$, which yields
\begin{align}\label{bessel2} \Gamma_\vektp = &
2\pi|\psi_{0}(\textbf{p})|^{2}\left(\frac{p^2}{2}+ \Ip\right)^{2}
\sum^{\infty}_{n=-\infty} \delta \Big(\frac{p^2}{2} + \Ip \\ & + \Up -
n \omega \Big) \Bigg[ J_{n}^{2} + \Xi_n(u,v) \sin \varphi
\Bigg]\nonumber
\end{align} where
\begin{align}\label{K} \Xi_n(u,v) = & - \frac{4 \xi J_{n}}{3} \Bigg[(4
v + n) ( J_{n-1} - J_{n+1}) \\ & - (J_{n-1} + J_{n+1} ) -
\frac{u}{8}(J_{n-2} - J_{n+2} ) \Bigg] . \nonumber
\end{align} For $\xi=0$, the known result \cite{Reiss1980} for the PES without the
$2\omega$ component is obtained. The rate \eqref{bessel2} has indeed
the same structure as the assumed yield in \eqref{yield}, that is
\begin{equation} \Gamma_\vektp = \Gamma_{0\vektp} + | \Delta
\Gamma_\vektp | \cos[\varphi+\Phi_1(\vektp)],
\end{equation} with the RPC $| \Delta \Gamma_\vektp |$ where
\begin{align}\label{rpcrate} \Delta \Gamma_\vektp = &
2\pi|\psi_{0}(\textbf{p})|^{2}\left(\frac{p^2}{2}+ \Ip\right)^{2} \\ &
\times \sum^{\infty}_{n=-\infty} \delta \Big(\frac{p^2}{2} + \Ip + \Up
- n \omega \Big) \Xi_n(u,v), \nonumber
\end{align} and the PP $\Phi_1(\vektp)=\pm \pi/2$, depending on the
sign of $\Delta \Gamma_\vektp$.  Evaluating the rate \eqref{bessel2}
 gives very similar results for PP and RPC as in
Fig.~\ref{fig:sfa:matrixelemRPCandPP}, as shown in
Fig.~\ref{fig:sfa:besselian}. The checkerboard structure is more pronounced in the analytical result  because of the flat-top, infinite-pulse assumption on which \eqref{bessel2} is based. For the numerical evaluation, the $\delta$ distribution in  \eqref{bessel2}  was replaced by a Gaussian $\exp[-(p^2/2+\Ip+\Up-n\omega)^2/a^2]/a\sqrt{\pi}$ with $a=1/20$, and the sum over $n$ was restricted to $\sum_{n=-30}^{30}$.

\section{Discussion} \label{disc}
The characteristic
checkerboard pattern in the momentum-resolved phase-of-the-phase spectra is
found in both theoretical and experimental results for xenon.
As mentioned above, this
pattern is related to above-threshold ionization peaks that are
shifted with varying relative phase. While the analytical calculation
based on the strong-field approximation predicts only
$\pm\sin\varphi$-like behavior of the yield for a $\sin\omega t +
\xi\sin(2\omega t + \varphi)$-like vector potential \eqref{vp}, the experiment
shows a more complex PP signature. Also, the experimental relative phase contrast is more oriented along the polarization axis than in the SFA. This clearly points towards the importance of Coulomb effects, which are expected to be the more pronounced the lower the photoelectron energy \cite{Bauer2006a,Popruzhenko2008}. 
Other
sources of discrepancy between SFA and experiment are due to focal
averaging, the projection onto the VMI detector plane in the measured PES \cite{vrakk2001}, possible resonance-enhanced multiphoton ionization (REMPI) \cite{Gong2017} and Freeman resonances \cite{Freeman91,Schyja98}. Comparison with PES from the numerical solution of the
time-dependent Schr\"odinger equation will be the subject of future
work. Preliminary TDSE results show that the PES also display
checkerboard patterns but also PP other than $\pm \sin$-like, as in the
experiment. However, achieving agreement in all details between TDSE and
experiment is challenging, even in the case of atomic hydrogen \cite{Kielp2014}.

\section{Summary and outlook} \label{concl}
An analytical expression for the ionization rate
in a two-color ($\omega$-$2\omega$), colinearly polarized laser field
was derived using the strong-field approximation. The change in the
yield as a function of the relative phase $\varphi$ between the two
color components was analyzed in the multiphoton regime using the
recently introduced phase-of-the-phase spectroscopy. A characteristic
checkerboard pattern in the momentum-resolved phase of the phase was
found in both theoretical and experimental results for xenon.

The visibility of the checkerboard pattern in the phase-of-the-phase spectra is
clearly related to the presence of above-threshold ionization peaks. Now, imagine an
experiment where photoelectron spectra from many-electron systems such as larger
molecules, clusters, fullerenes, droplets, nanospheres, or solids are
measured. Because of the various relaxation channels, above-threshold ionization peaks or
other patterns in photoelectron spectra based on the interference of different electron
pathways may be masked by delayed or even thermal electron emission,
leading to insipidly Maxwellian-like spectra. Phase-of-the-phase
spectroscopy effectively removes these incoherently emitted electrons
from the spectrum because incoherent emission is independent of the
relative phase. This idea was applied to the electron emission from
SiO$_2$ nanospheres due to the interaction with strong, few-cycle
laser fields \cite{Zhereb2012} where the phase-of-phase analysis was performed with respect to
the carrier-envelope frequency instead of the relative phase between
the two colors in a two-color laser field. Clearly, phase-of-the-phase spectroscopy
can be applied with respect to any periodic ``knob'' that can be controlled in an
experiment.

\section*{Acknowledgment} This work was supported by the SFB 652 and
projects BA 2190/8 and TI 210/7 of the German Science Foundation (DFG).

\bibliography{bibliography}

\end{document}